\documentclass[two column,letter]{revtex4-1}
\usepackage{color}
\usepackage{graphicx}
\begin{document}

\title{Evidence for $\Gamma_{8}$ Ground-State Symmetry of Cubic YbB$_{12}$\\ Probed by Linear Dichroism in Core-Level Photoemission}
\author{Yuina Kanai$^{1, 2}$, Takeo Mori$^{1}$, Sho Naimen$^{1}$, Kohei Yamagami$^{1, 2}$, Hidenori Fujiwara$^{1, 2}$,\\ Atsushi Higashiya$^{2, 3}$, Toshiharu Kadono$^{2, 4}$, Shin Imada$^{2, 4}$, Takayuki Kiss$^{1, 2, 5}$,\\ Arata Tanaka$^{6}$, Kenji Tamasaku$^{2, 5}$, Makina Yabashi$^{2}$,\\ Tetsuya Ishikawa$^{2}$, Fumitoshi Iga$^{7}$, and Akira Sekiyama$^{1, 2, 5}$}
\affiliation{$^1$Division of Materials Physics, Graduate School of Engineering Science, Osaka University, Toyonaka, Osaka 560-8531, Japan \\
$^2$RIKEN SPring-8 Center, Sayo, Hyogo 679-5148, Japan \\
$^3$Faculty of Science and Engineering, Setsunan University, Neyagawa, Osaka 572-8508, Japan \\
$^4$Department of Physical Sciences, Ritsumeikan University, Kusatsu, Shiga 525-8577, Japan \\
$^5$Center for Promotion of Advanced and Interdisciplinary Research, Graduate School of 
Engineering Science, Osaka University, Toyonaka, Osaka 560-8531, Japan \\ 
$^6$Department of Quantum Matter, ADSM, Hiroshima University, Higashi-Hiroshima, Hiroshima 739-8530, Japan \\
$^7$College of Science, Ibaraki University, Mito, Ibaraki 310-8512, Japan\\}
\begin{abstract}
We have successfully observed linear dichroism in angle-resolved Yb$^{3+}$ 3$d_{5/2}$ core-level photoemission spectra for YbB$_{12}$ in cubic symmetry. Its anisotropic 4$f$ charge distribution due to the crystal-field splitting is responsible for the linear dichroism, which has been verified by spectral simulations using ionic calculations with the full multiplet theory for a single-site Yb$^{3+}$ ion in cubic symmetry. The observed linear dichroism as well as the polarization-dependent spectra in two different photoelectron directions for YbB$_{12}$ are quantitatively reproduced by theoretical analysis for the $\Gamma_{8}$ ground state, indicating the $\Gamma_{8}$ ground-state symmetry for the Yb$^{3+}$ ions mixed with the Yb$^{2+}$ state.
\end{abstract}
\maketitle

Rare-earth-based strongly correlated electron systems show various interesting phenomena such as competition between magnetism and unconventional superconductivity, charge and/or multipole ordering, and the formation of a narrow ($\sim$meV) gap at low temperatures. 
Among them, YbB$_{12}$ is known as a Kondo semiconductor~\cite{Kasaya1983,Kasaya1985,Iga98,Susaki99,SasoHarima}, which has been recently recognized as a candidate for topological insulators~\cite{YbB12topo}, as intensively discussed for another Kondo semiconductor, SmB$_{6}$~\cite{SmB6Suga,SmB6Takimoto,topoKondoInsulator}. 
The mean valence of YbB$_{12}$ has been estimated as $\sim$2.9+ by bulk-sensitive 3$d$ core-level hard X-ray photoemission (HAXPES) spectroscopy~\cite{YbB12PRB2009}. 
To discuss the mechanisms of the gap opening at low temperatures~\cite{SasoHarima,JPCM1995,PSRPRB2003,PRL1995,PRB2002,PRB2009,Yamaguchi2013} and the possibility of a topological insulator, it is essential to verify the ground-state symmetry of the Yb$^{3+}$ [4$f^{13}$ (one hole)] state determined by crystalline-electric-field (CEF) splitting. 
Nevertheless, it is unclear for YbB$_{12}$. 
Its eightfold degenerated Yb$^{3+}$ 4$f_{7/2}$ levels are considered to split into two quartets~\cite{YbB12andCeB6} (normally, two doublets and a quartet) due to CEF in YbB$_{12}$. 
The ground state of the Yb$^{3+}$ ions is commonly asserted to be in the so-called $\Gamma_{8}$ symmetry~\cite{SasoHarima,YbB12neutron},  but a possible accidentally degenerated $\Gamma_{6} + \Gamma_{7}$ ground state is not completely excluded~\cite{YbB12andCeB6}. 
In this Letter, we show evidence for the $\Gamma_{8}$ symmetry of the Yb$^{3+}$ sites in the ground state of YbB$_{12}$ on the basis of linear dichroism in {\it angle-resolved} core-level photoemission.

Generally, it is difficult to experimentally determine the 4$f$ ground-state symmetry. 
Inelastic neutron scattering is useful, but other excitations such as phonon excitations often hamper the observation of magnetic 4$f-$4$f$ excitations. 
Linear dichroism (LD) in 3$d-$to$-$4$f$ soft X-ray absorption spectroscopy (XAS) for single crystals is powerful owing to the dipole selection rules, as reported for Ce compounds~\cite{WillersCePt3Si,Hansmann08LD,WillersCeTIn5,WillersCe122}. 
However, it is not applicable to compounds in cubic symmetry, in which there is no anisotropic axis relative to the electric field of the incident light. On the other hand, the selection rules work also in the photoemission process while the excited electron energy is much higher than that in the absorption. 
Furthermore, there is another controllable measurement parameter in photoemission, called the  ``photoelectron detection direction" relative to the single-crystalline axis, in addition to the polarization direction of the excitation light. 
Indeed, by using LD in 3$d$ core-level HAXPES spectra, the Yb$^{3+}$ 4$f$ ground state has been determined for tetragonal YbCu$_{2}$Si$_{2}$ and YbRh$_{2}$Si$_{2}$~\cite{MoriYCSYRS}. LD in the core-level HAXPES for cubic Yb compounds is also expected to be observed, as discussed below.

\begin{figure}
\begin{center}
\includegraphics[width=8.5cm]{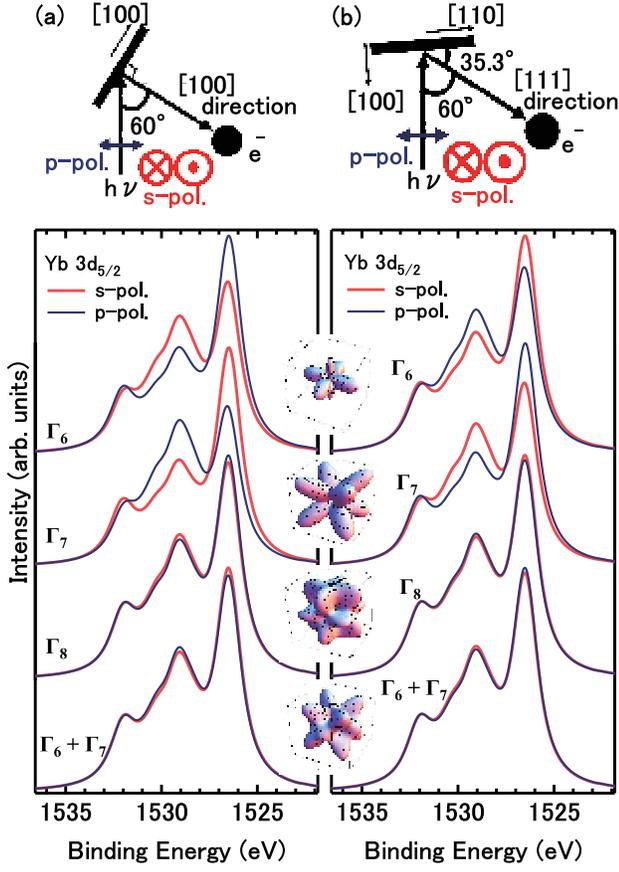}
\end{center}
\vspace{2mm}
\begin{center}
\caption{(Color online) (a) Simulated polarization-dependent 3$d_{5/2}$ photoemission spectra 
of Yb$^{3+}$ ions assuming the crystal-field-split ground state in cubic symmetry in the [100] direction, 
together with the corresponding experimental geometry. (b) Same as (a) but for the photoelectron in the [111] direction. The 4$f$-hole spatial distributions for the corresponding states are also shown.}
\label{Simulations}
\end{center}
\end{figure}

In the case of Yb$^{3+}$ ions in tetragonal symmetry, the eightfold degenerate $J=7/2$ state splits into four doublets as
\begin{eqnarray}
&&|\Gamma_7^1\rangle = c|\pm5/2\rangle+\sqrt{1-c^2}|\mp3/2\rangle,\label{G71} \\
&&|\Gamma_7^2\rangle =- \sqrt{1-c^2}|\pm5/2\rangle+c|\mp3/2\rangle,\label{G72} \\
&&|\Gamma_6^1\rangle = b|\pm1/2\rangle+\sqrt{1-b^2}|\mp7/2\rangle,\label{G61}\\
&&|\Gamma_6^2\rangle =\sqrt{1-b^2}|\pm1/2\rangle-b|\mp7/2\rangle,\label{G62}
\end{eqnarray}
where the coefficients $0\leq b\leq 1, 0\leq c\leq 1$ defining the actual charge distributions 
and CEF splitting energies depend on the CEF parameters 
$B_2^0, B_4^0, B_4^4, B_6^0$, and $B_6^4$ in Stevens formalism~\cite{Stevens}. In the cubic symmetry, the eightfold degenerate $J=7/2$ state splits generally into two doublets and one quartet as
\begin{eqnarray}
&&|\Gamma_6\rangle = \sqrt{5/12}|\pm7/2\rangle+\sqrt{7/12}|\mp1/2\rangle,\label{G6} \\
&&|\Gamma_7\rangle =- \sqrt{3}/2|\pm5/2\rangle+1/2|\mp3/2\rangle,\label{G7} \\
&&|\Gamma_8\rangle = 
\left\{
\begin{array}{l}
-\sqrt{7/12}|\pm7/2\rangle+\sqrt{5/12}|\mp1/2\rangle\\
1/2|\pm5/2\rangle+\sqrt{3}/2|\mp3/2\rangle
\end{array}
\right..\label{G8}
\end{eqnarray}
The $\Gamma_{6}$ and $\Gamma_{7}$ states correspond to the tetragonal $\Gamma_{6}^{1}$ (with $b = \sqrt{7/12}$) and $\Gamma_{7}^{2}$ (with $c = 1/2$) states, respectively. 
Since their 4$f$ charge distributions deviate from spherical symmetry owing to the CEF splitting even in the cubic symmetry, as shown in Fig.~\ref{Simulations}, it is natural to expect the observation of LD in core-level photoemission for cubic Yb compounds. 
The $\Gamma_{6}$, $\Gamma_{7}$, and $\Gamma_{8}$ 4$f$ charge distributions are elongated along the [100], [111], and [110] directions, respectively. Actually, we have performed ionic calculations including the full multiplet 
theory~\cite{Thole85} and the local CEF splitting using the XTLS 9.0 program~\cite{XTLS}. 
All atomic parameters such as the 4$f$-4$f$ and 3$d$-4$f$ Coulomb and exchange 
interactions (Slater integrals) and the 3$d$ and 4$f$ spin-orbit couplings have been 
obtained using Cowan's code~\cite{Cowan} based on the Hartree-Fock method. 
The Slater integrals (spin-orbit couplings) are reduced to 88\% (98\%) to fit 
the core-level photoemission spectra~\cite{YbB12PRB2009}.

Simulated polarization-dependent Yb$^{3+}$ $3d_{5/2}$ core-level HAXPES spectra in cubic symmetry at two photoelectron directions ([100] and [111]) are shown in Fig.~\ref{Simulations}. 
In the case of YbB$_{12}$, since the accidentally degenerated $\Gamma_{6} + \Gamma_{7}$ states could be a candidate for the ground state, the simulations  assuming the $\Gamma_{6} + \Gamma_{7}$ ground state and its $4f$ hole spatial distribution are also shown. 
LD defined by the difference in spectral weight between the s- and p-polarization configurations is reversed between the [100] and [111] directions for all states displayed here. 
LD for the $\Gamma_{6}$ ground state has the same tendency as that for the $\Gamma_{8}$ ground state. 
On the other hand, LD for the $\Gamma_{8}$ state is the smallest since the 4$f$ hole spatial distribution for the $\Gamma_{8}$ state is the nearest to a spherical shape among these three eigenfunctions. LD assuming the $\Gamma_{6} + \Gamma_{7}$ state is reversed to that for the $\Gamma_{8}$ state. These simulations indicate that the symmetry of the Yb$^{3+}$ state can be determined by LD in the core-level photoemission.

We have performed LD in HAXPES~\cite{SekiyamaAuAg,ASHAXPES2013} 
at BL19LXU of SPring-8~\cite{YabashiPRL01}  
using a MBS A1-HE hemispherical photoelectron spectrometer. 
A Si(111) double-crystal monochromator selected linearly polarized 7.9 keV radiation 
within the horizontal plane, 
which was further monochromatized using a Si(620) channel-cut crystal. 
To switch the linear polarization of the excitation light from the horizontal direction 
to the vertical direction, two single-crystalline (100) diamonds were used as a phase retarder  placed downstream of the channel-cut crystal. 
The $P_L$ (degree of linear polarization) of the polarization-switched X-ray after the phase retarder was estimated as 
$-0.93$, corresponding to the vertically linear polarization component of 96.5\%.  
Since the detection direction of 
photoelectrons was set in the horizontal plane 
with an angle to incident photons of 60$^{\circ}$, as shown in Figs.~\ref{Simulations}(a) and \ref{Simulations}(b),  
the experimental configuration 
at the horizontally (vertically) polarized light excitation corresponds to the p-polarization 
(s-polarization). 
The excitation light was focused onto the samples using an ellipsoidal Kirkpatrik-Baez mirror. 
To precisely detect LD in the Yb 3$d$ core-level photoemission spectra, we optimized the photon flux so as to set comparable photoelectron count rates between the s- and p-polarization configurations.
Single crystals of YbB$_{12}$ 
synthesized by the traveling-solvent floating-zone method~\cite{Iga98,Susaki99} were fractured along the (100) plane {\it in situ}, 
where the base pressure was $\sim$1$\times10^{-7}$ Pa. The experimental geometry was controlled using a newly developed two-axis manipulator~\cite{H.Fujiwara}, where the normal emission direction parallel to the [100] direction in Fig.~\ref{Simulations}(a) was changed to the photoelectron detection in the [111] direction in Fig.~\ref{Simulations}(b) by azimuthal rotation of 45$^\circ$ and polar rotation by $\sim$55$^\circ$.
The sample and surface qualities were examined on the basis of the absence of 
any core-level spectral weight caused by possible impurities including oxygen and carbon. 
The energy resolution was set to 400 meV. 
The measuring temperature is 9 K, which is sufficiently lower than the excited state ($\gtrsim250$ K)~\cite{YbB12neutron}.

\begin{figure}
\begin{center}
\includegraphics[width=8.5cm]{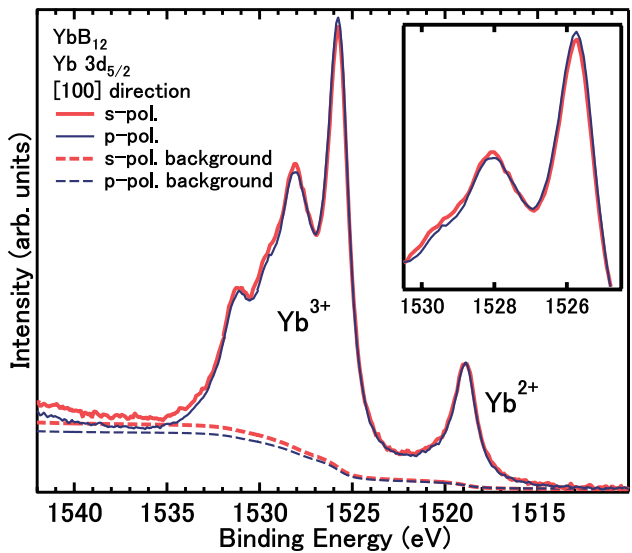}
\end{center}
\vspace{2mm}
\begin{center}
\caption{(Color online) Polarization-dependent Yb 3$d_{5/2}$ core-level HAXPES raw spectra (solid lines)
of YbB$_{12}$ in the [100] direction and optimized Shirley-type backgrounds (see text), which we have subtracted from the raw spectra (dashed lines). 
The raw spectra have been normalized by the Yb$^{2+}$ spectral weight. 
The same spectra in the expanded scale are also shown.}
\label{RawData}
\end{center}
\end{figure}

The polarization-dependent Yb$^{3+}$ 3$d_{5/2}$ HAXPES spectra in the [100] direction of 
YbB$_{12}$ are shown in Fig.~\ref{RawData}. 
A single peak at a binding energy of $\sim$1519 eV and a multiple peak ranging from 1524 to 1534 eV exist in all the spectra. Since the 4$f$ subshell is fully occupied in the Yb$^{2+}$ sites with a spherically symmetric 4$f$ distribution, 
the former single peak is ascribed to the Yb$^{2+}$ states. 
The $3d^94f^{13}$ final states 
for the Yb$^{3+}$ components show an atomic-like multiplet-split peak structure 
in the $1524-1534$ eV range. We show the same raw spectra in the expanded scale ranging from $1524.5$ to $1530.5$ eV in Fig.~\ref{RawData}. 
A slight but intrinsic LD is seen in the highest and second-highest peaks in the raw spectra. 
The so-called Shirley-type backgrounds are also displayed in Fig.~\ref{RawData}. 
We have optimized the backgrounds as follows: 
After the normalization of the background-subtracted spectra by both Yb$^{2+}$ and Yb$^{3+}$ $3d_{5/2}$ spectral weights, the relative Yb$^{2+}/$Yb$^{3+}$ contributions and the intensities in the high-binding-energy region of $1534-1540$ eV become equivalent between the s- and p-polarization configurations. 
The reference binding energy on the higher side has been set to 1536.2 eV corresponding to the local minimum of the raw spectral weight in the p-polarization. 
As a result, there are finite spectral weights at $\sim$1536 eV in the background-subtracted spectra. 
However, these should be intrinsic owing to the overlap of the tails of the lifetime-broadened Yb$^{3+}$ $3d_{5/2}$ main peaks and a plasmonic energy-loss structure at the higher binding energy. 
Note that we have confirmed the robustness of LD in the $1524-1531$ eV region regardless of background intensity.

\begin{figure}
\begin{center}
\includegraphics[width=8.5cm]{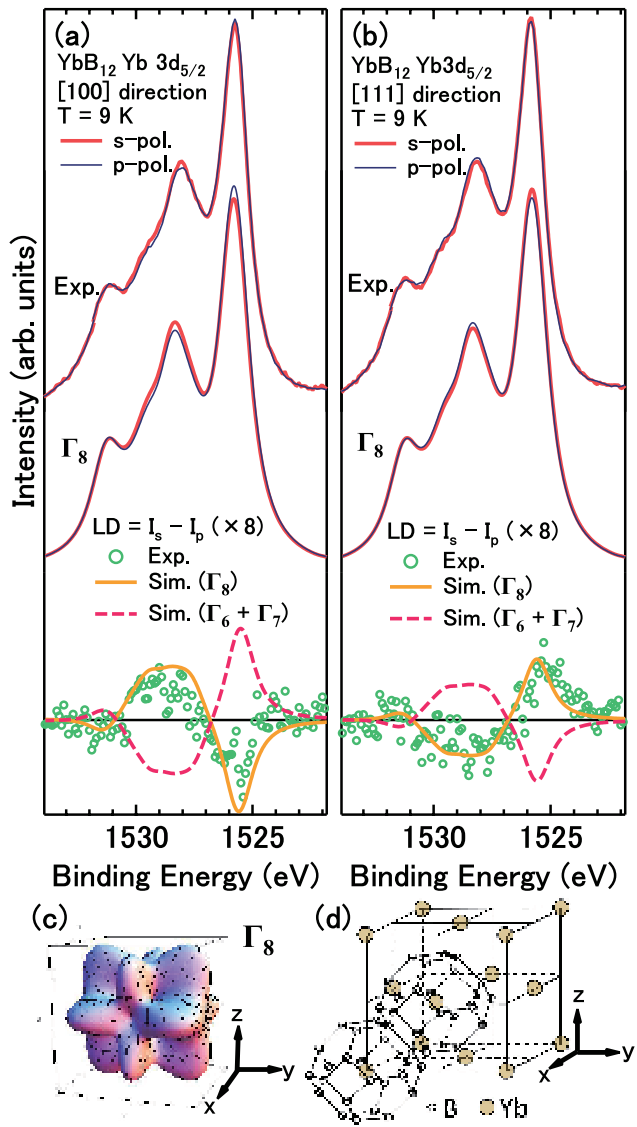}
\end{center}
\vspace{2mm}
\begin{center}
\caption{(Color online) (a) Polarization-dependent Yb$^{3+}$ 3$d_{5/2}$ core-level HAXPES spectra and LD of YbB$_{12}$ compared with the simulated ones for the $\Gamma_{8}$ ground state in the [100] direction. 
Simulated LD assuming the accidentally degenerated $\Gamma_{6} + \Gamma_{7}$ ground state is also shown by the dashed line in the lower panel. 
(b) Same as (a) but data in the [111] direction. 
The Shirley-type background has been subtracted from the raw spectra (Fig.~\ref{RawData}). (c) 4$f$-hole spatial distribution for the Yb$^{3+}$ ion with the  $\Gamma_{8}$ symmetry. (d) Crystal structure of YbB$_{12}$~\cite{UB12crystalstructure}}
\label{Results}
\end{center}
\end{figure}

A comparison of the polarization-dependent background-subtracted Yb$^{3+}$ 3$d_{5/2}$ HAXPES spectra of YbB$_{12}$ and their LD with the photoelectron directions of [100] and [111] with the simulated ones for the $\Gamma_{8}$ ground state is shown in Fig.~\ref{Results}. 
The highest peak is slightly stronger in the p-polarization configuration (p-pol.) than in the s-polarization one (s-pol.), and the second highest peak is stronger in the s-pol. for both experimental and simulated spectra in the [100] direction. 
These tendencies are reversed in the data in the [111] direction. 
As shown in Figs.~\ref{Results}(a) and \ref{Results}(b), the observed LD and spectra are quantitatively reproduced by the simulations for the $\Gamma_{8}$ ground state, for which the 4$f$ charge distribution is shown in Fig.~\ref{Results}(c). 
If the 4$f$ ground state of YbB$_{12}$ were in the $\Gamma_{6}$ symmetry, LD would be much larger than the experimental one. 
The sign of LD for the $\Gamma_{7}$ or accidentally degenerated $\Gamma_{6} + \Gamma_{7}$ ground state is completely inconsistent with our experimental results. 
Such a quantitative reproducibility of the observed LD and core-level spectra by the simulations surely indicates the Yb$^{3+}$ (4$f^{13}$) ions in the $\Gamma_{8}$ symmetry mixed with a small quantity ($\sim10\%$) of the Yb$^{2+}$ (4$f^{14}$) component~\cite{YbB12PRB2009} in the ground state of YbB$_{12}$.

Our finding of the $\Gamma_{8}$ 4$f$ symmetry for the Yb$^{3+}$ sites in the ground state is consistent with the prediction by a band-structure (local density approximation $+$ on-site Coulomb repulsion, LDA$+U$) calculation, where the $\Gamma_{6}$ and $\Gamma_{7}$ states are on the occupied side~\cite{SasoHarima}. 
One might consider that the $\Gamma_{6}$ and $\Gamma_{7}$ states are possibly mixed in the ground state due to the hybridization between the 4$f$ and valence-band orbitals at low temperatures well below the Kondo temperature ($\sim240$ K for YbB$_{12}$~\cite{Kasaya1983,Kasaya1985,Iga98,Susaki99}). 
As widely recognized for the local electronic structures of transition metal oxides, on the other hand, 
note that the so-called CEF splitting actually seen in realistic materials is a consequence of the anisotropic hybridization effects in addition to static ligand potentials. 
The latter would be much smaller than the former for YbB$_{12}$ since the ligand-field potential on the Yb sites is to some extent nearer to a spherically symmetric one, being  caused by the crystal structure~\cite{Kasaya1983,Kasaya1985,UB12crystalstructure} where the Yb ion is surrounded by the truncated octahedron made of 24 boron ions as shown in Fig.~\ref{Results}(d). 
The 4$f$-hole spatial distributions are elongated along the centers of the truncated octahedron faces for the $\Gamma_{6}$ and $\Gamma_{7}$ states, whereas they are elongated along the edges of the hexagonal faces for the $\Gamma_{8}$ state, 
which leads to the conclusion that the 4$f$ holes in the $\Gamma_{8}$ state are relatively stabilized by the hybridizations compared with those in the $\Gamma_{6}$ and $\Gamma_{7}$ states with different symmetries, as suggested by the LDA$+U$ calculation. 
Then, the possible $\Gamma_{6}$ and/or $\Gamma_{7}$ state mixture would be experimentally insignificant in our data.

In summary, we have successfully determined the 4$f$ symmetry of the Yb$^{3+}$ sites in the ground state for cubic YbB$_{12}$ as the $\Gamma_{8}$ symmetry by LD in the Yb$^{3+}$  core-level HAXPES in two different photoelectron directions. 
Our result also suggests that the Yb$^{3+}$ ion model under the effective CEF, in which the hybridization effects are implicitly taken into account, is suitable even for the valence-fluctuating system at low temperatures well below the Kondo temperature. 
The applicability of LD in the core-level HAXPES even to the system in the cubic symmetry (not restricted to systems with lower symmetry) demonstrated here would be promising for revealing the strongly correlated orbital symmetry of the ground state in a partially filled subshell, 
the charge distribution of which deviates from the spherical symmetry. 

\acknowledgments{We thank H. Yomosa, S. Fujioka, K. Yano, Y. Nakata, Y. Nakatani, T. Yagi, S. Tachibana, Y. Nakamura, H. Aratani, and K. Kodera for supporting the experiments. We are also grateful to K. Miyake, A. Tsuruta, Y. Saitoh, A. Yasui, and A. Fujimori for fruitful discussions. This work was supported by a Grant-in-Aid for Scientific Research
(23654121), 
Grants-in-Aid for Young Scientists (23684027 and 23740240),
and a Grant-in-Aid for Innovative Areas (20102003) from MEXT and JSPS,
Japan, 
and by Toray Science Foundation.
The hard X-ray photoemission was performed at SPring-8 under the approval of
JASRI (2014A1149, 2014B1305). }

\end{document}